\documentclass[pra,amsmath,amssymb,twocolumn,superscriptaddress]{revtex4}

\usepackage{placeins}
\usepackage{float}
\usepackage{subfigure}

\usepackage{amsmath,amssymb}
\usepackage[usenames]{color}
\usepackage{amssymb}
\usepackage{grffile}
\usepackage{graphicx}
\usepackage{amsmath, amstext, amssymb, amsfonts, amsxtra}
\usepackage{textcomp}
\usepackage{xspace}
\usepackage{bbm}
\usepackage{bm}
\newcommand{\be}{\begin{equation}}
\newcommand{\ee}{\end{equation}}

\input{epsf}

\newcommand{\rem}[1]{}

\begin{document}

\title{Quantum computation of complex systems}
\author{Giuliano Benenti}
\email{giuliano.benenti@uninsubria.it}
\affiliation{Center for Nonlinear and Complex Systems, Dipartimento
di Scienza e Alta Tecnologia,
Universit\`a degli Studi dell'Insubria, via Valleggio 11, 22100 Como, Italy}
\affiliation{Istituto Nazionale di Fisica Nucleare, Sezione di Milano,
via Celoria 16, 20133 Milano, Italy}
\author{Giulio Casati}
\email{giulio.casati@uninsubria.it}
\affiliation{Center for Nonlinear and Complex Systems, Dipartimento
di Scienza e Alta Tecnologia,
Universit\`a degli Studi dell'Insubria, via Valleggio 11, 22100 Como, Italy}
\date{\today}

\maketitle

\section*{Key points}

\begin{itemize}

\item
Quantum logic: superposition and entanglement

\item
Quantum algorithms for complex dynamical systems

\item
Quantum simulations on actual quantum hardware

\end{itemize}

\section{Introduction}

Miniaturization provides us with an intuitive way of
understanding why, in the near future, quantum mechanics will become
important for computation. The electronics industry
for computers grows hand-in-hand with the decrease in size of integrated 
circuits. This miniaturization is necessary to increase 
computational power, that is, the
number of floating-point operations per second (flops) a
computer can perform. In the 1950's, electronic computers based on vacuum-tube
technology were capable of performing approximately $10^3$ floating-point 
operations per second, while nowadays (2022) there exist
supercomputers whose power is greater than $100$ petaflops 
(a 1 petaflops computer is capable of performing 
$10^{15}$ floating-point operations per second).
This enormous growth of computational power
has been made possible owing to progress in
miniaturization, which may be quantified 
empirically in Moore's law. This law is the result of a remarkable
observation made by Gordon Moore in 1965: the number of
transistors on a single integrated-circuit chip doubles approximately
every $18-24$ months. This exponential growth has not yet
saturated and Moore's law is still valid. At the present
time the limit is close to $10^{10}$ transistors per
chip and the typical size of circuit components is of the order
of $5-10$ nanometres. Extrapolating Moore's
law, it is estimated that within a few years, one
would reach the atomic size for storing a single bit of
information. At that point, quantum effects will become
unavoidably dominant.

Quantum physics sets fundamental limitations on
the size of the circuit components. 
The first question under debate is
whether it would be more convenient to push the silicon-based transistor
to its physical limits or instead to
develop alternative devices, such as quantum dots,
single-electron transistors or molecular switches. A
common feature of all these devices is that they are at the nanometre
length scale, and therefore quantum effects play a crucial role.

So far, the quantum switches that could substitute
silicon-based transistors and possibly be connected together to execute
classical algorithms based on Boolean logic were discussed. In this perspective, quantum
effects are simply unavoidable corrections that must be taken into
account owing to the nanometre size of the switches. A
quantum computer represents a radically different challenge: the aim is to
build a machine based on quantum logic, that is, it
processes the information and performs logic operations in agreement with the
laws of quantum mechanics~\cite{qcbook}.

\section{Quantum logic}

The elementary unit of quantum information is called a qubit (the quantum
counterpart of the classical bit) and a quantum computer may be
viewed as a many-qubit system. Physically, a qubit is a
two-level system, like the two spin states of a spin-$1/2$
particle, the vertical and horizontal polarization states of a single
photon or two levels of an atom. 
 
A classical bit is a system that can exist in two distinct states, which are
used to represent $0$ and $1$, that is, a single binary digit. The only
possible operations (gates) in such a system are the identity
($0\to0$, $1\to1$) and NOT ($0\to1$, $1\to0$). 
In contrast, a quantum bit (qubit) is a two-level quantum system,
described by a two-dimensional complex Hilbert space. In this space, one
may choose a pair of normalized and mutually orthogonal quantum
states, called $|0\rangle$ and $|1\rangle$ (say, the eigenstates of the Pauli 
operator $\sigma_z$), to represent the values $0$ and $1$ of a 
classical bit. These two states form a computational basis.
From the superposition principle, any state of the qubit 
may be written as
\begin{equation}
  |\psi\rangle = \alpha |0\rangle + \beta |1\rangle \,,
  \label{alphabetaqubit}
\end{equation}
where the amplitudes $\alpha$ and $\beta$ are complex numbers,
constrained by the normalization condition
$|\alpha|^2 + |\beta|^2 = 1$.

A quantum computer can be seen as a collection of $n$
qubits and therefore its wave function resides in a $2^n$-dimensional
complex Hilbert space.
While the state of an $n$-bit classical computer is described in binary
notation by an integer $k\in [0,2^n-1]$,
\begin{equation}
  k = k_{n-1} \, 2^{n-1} + \cdots + k_1 \, 2 + k_0 \,,
  \label{binary}
\end{equation}
with $k_0,k_1,\dots ,k_{n-1}\in [0,1]$ binary digits, the state of an
$n$-qubit quantum computer is
\begin{align}
  |\psi\rangle & =
  \sum_{k=0}^{2^n - 1} c_k \, |k\rangle
  \nonumber
  \\ &=
  \sum_{k_{n-1},...,k_1,k_0=0}^1
  c_{k_{n-1}, \dots, k_1, k_0 } \,
  |k_{n-1}\cdots k_1 k_0\rangle,
  \label{superposition}
\end{align}
where $|k_{n-1}\cdots k_1 k_0\rangle\equiv |k_{n-1}\rangle \otimes \cdots
\otimes |k_1\rangle \otimes |k_0\rangle$. Notice that the 
complex numbers $c_k$ are constrained by the normalization condition
$\sum_{k=0}^{2^n-1} |c_k|^2 = 1$.

The superposition principle is clearly visible in
Eq.~(\ref{superposition}): while $n$ classical bits can store only a single
integer $k$, the $n$-qubit quantum register can be prepared in the
corresponding state $|k\rangle$ of the computational basis, but also in a
superposition. The number of states of the computational basis
in this superposition can be as large as $2^n$, which grows
exponentially with the number of qubits. The superposition principle opens up
new possibilities for computation. When one performs 
a computation on a classical
computer, different inputs require separate runs. In contrast, a
quantum computer can perform a computation for exponentially many inputs on a
single run. This huge parallelism is the basis of the power of quantum
computation.

The superposition principle is not a uniquely quantum feature.
Indeed, classical waves satisfying the superposition
principle do exist. For instance, consider the wave
equation for a vibrating string with fixed endpoints. Its solutions
$|\varphi_k\rangle$ satisfy the superposition principle and one can write
the most general state $|\varphi\rangle$ of a vibrating string as a linear
superposition of these solutions, analogously to Eq.~(\ref{superposition}):
$|\varphi \rangle = \sum_{k=0}^{2^n-1} c_{k} |\varphi_k\rangle$.
It is therefore also important to point out the importance of
entanglement for the power of quantum computation,
as compared to any classical computation.
Entanglement is the most spectacular and counter-intuitive
manifestation of quantum mechanics, observed in composite quantum systems: it
signifies the existence of non-local correlations between
measurements performed on well-separated particles. After
two classical systems have interacted, they are in
well-defined individual states. In contrast, after
two quantum particles have interacted, in general, they can
no longer be described independently of each other. There
will be purely quantum correlations between two
such particles, independently of their spatial separation.
Examples of two-qubit entangled state are the four states of the 
so-called Bell basis, 
$|\phi^{\pm}\rangle=\frac{1}{2}(|00\rangle\pm|11\rangle)$ and 
$|\psi^{\pm}\rangle=\frac{1}{2}(|01\rangle\pm|10\rangle)$.
The measure of the polarization state of one qubit will 
instantaneously affect 
the state of the other qubit, whatever their distance is. 
There is no entanglement in classical physics. 
Therefore, in order to represent the superposition of $N=2^n$ levels
by means of classical waves, these levels must belong to the same system.
Indeed, classical states of separate systems can never be superposed.
Thus, any computation based on classical waves
requires a number $N$ of levels that
grows exponentially with $n$. If $\Delta$ is the typical energy separation
between two consecutive levels, the amount of energy required for this
computation is given by $\Delta2^n$. Hence, the amount of physical
resources needed for the computation grows exponentially with $n$.
In contrast, due to entanglement, in quantum
physics a general superposition of $2^n$ levels may be represented by means of
$n$ qubits. Thus, the amount of physical resources (energy) grows only
linearly with $n$.

To implement a quantum computation, one must be able to control the 
evolution in time of the many-qubit state describing the quantum
computer. As far as the coupling to the
environment may be neglected, this evolution is unitary and
governed by the Schr{\"o}dinger equation. 
It is well known that a small set of elementary logic gates
allows the implementation of any complex computation
on a classical computer. This is very important: it means that, when one 
changes
the problem, one does not need to modify one's computer hardware. Fortunately, 
the same property remains valid for a quantum computer. 
It turns out that, 
in the quantum circuit model, any whatever complex
unitary transformation acting on 
a many-qubit system can be decomposed into quantum gates acting 
on a single qubit and a suitable quantum gate acting on two qubits.
Any unitary operation on a single qubit can be
constructed using only Hadamard and phase-shift gates. 
The Hadamard gate is defined as follows: it turns 
$|0\rangle$ into $(|0\rangle + |1\rangle)/\sqrt{2}$ and  
$|1\rangle$ into $(|0\rangle - |1\rangle)/\sqrt{2}$.
The phase-shift gate (of phase $\delta$) turns 
$|0\rangle$ into $|0\rangle$ and $|1\rangle$ into
$e^{i\delta}|1\rangle$. 
One can decompose a generic unitary transformation acting on 
a many-qubit state 
into a sequence of Hadamard, phase-shift and 
CNOT gates, where CNOT is a two-qubit gate, defined as follows:
it turns $|00\rangle$ into $|00\rangle$, 
$|01\rangle$ into $|01\rangle$, 
$|10\rangle$ into $|11\rangle$, and 
$|11\rangle$ into $|10\rangle$.
As in the classical XOR gate, the CNOT gate flips the state of the 
second (target) qubit if the first (control) qubit is in the state 
$|1\rangle$ and does nothing if the first qubit is in the state 
$|0\rangle$.
Of course, the CNOT gate, in contrast to the
classical XOR gate, can also be applied to any superposition of the
computational basis states. 

The decomposition of a generic unitary transformation of a $n$-qubit 
system into elementary quantum gates is in general inefficient,
that is, it requires a number of gates exponentially large 
in $n$ (more precisely, $O(n^2 4^n)$ quantum gates).
However, there are special unitary transformations that can
be computed efficiently in the quantum circuit model, namely
by means of a number of elementary gates polynomial in $n$. 
A very important example is given by the quantum Fourier transform,
mapping a generic $n$-qubit state $\sum_{k=0}^{2^n-1} a_k |k\rangle$ 
into $\sum_{l=0}^{2^n-1} b_l |l\rangle$, where the vector 
$\{b_0,...,b_{N-1}\}$ is the discrete Fourier transform of the vector 
$\{a_0,...,a_{N-1}\}$, that is, 
$b_l=\sum_{k=0}^{N-1} e^{2\pi i k l/ 2^n} a_k$. It can be shown that
this transformation can be efficiently implemented in 
$O(n^2)$ elementary quantum gates, whereas the best known classical 
algorithm to simulate the Fourier transform, the fast Fourier transform, 
requires $O(n 2^n)$ elementary operations. 
The quantum Fourier transform is an essential subroutine in many quantum 
algorithms.

\section{Quantum algorithms}

As shown above, the power of quantum computation is due to the inherent
quantum parallelism associated with the superposition principle.
In simple terms, a quantum computer can process a large number of 
classical inputs in a single run.
For instance, starting from the input state 
$\sum_{k=0}^{2^n-1}c_k|k\rangle \otimes |0...0\rangle$, 
one may obtain the output 
state 
\begin{equation}
\sum_{k=0}^{2^n-1}c_k|k\rangle \otimes |f(k)\rangle.
\label{fsuperposition}
\end{equation}
Therefore, the function $f(k)$ is computed for all $k$ in a 
single run (note that one needs two quantum registers to compute
by means of a reversible unitary transformation $f(k)$; the second 
register requires enough qubits to load the output $f(k)$ for all input values
$k=0,1,..,2^n$, with $n$ number of qubits in the first register). 
However, it is not an easy task to extract useful information 
from the output state. The problem is that this information
is, in a sense, hidden.
Any quantum computation ends up with a projective measurement
in the computational basis: the
qubit polarization is measured along the $z$-axis for all the qubits. 
The output of the
measurement process is inherently probabilistic and the probabilities of
the different possible outputs are set by the basic postulates of quantum
mechanics. Given the state (\ref{fsuperposition}), one obtains 
$|\bar{k}\rangle|f(\bar{k})\rangle$ with probability $|c_{\bar{k}}|^2$,
hence, the evaluation of the function $f(k)$ for a single 
$k=\bar{k}$, exactly as with a classical computer. However, there exist
quantum algorithms that exploit quantum
interference to efficiently extract useful information.

In 1994, Peter Shor proposed a quantum algorithm that efficiently solves
the prime-factorization problem: given a composite odd positive 
integer $N$, find its prime factors. 
This is a central problem in computer science and it is
conjectured, though not proven, that for a classical computer it
is computationally difficult to find the prime factors. 
Indeed, the best classical 
algorithm, the number field sieve, requires
$\exp(O(n^{1/3}(\log{n})^{2/3}))$ operations.
Shor's
algorithm instead efficiently solves the integer factorization problem 
in $O((n^2\log{n}\log\log{n}))$ elementary quantum gates, where 
$n=\log N$ is the number of bits necessary to code the input $N$.
Therefore it provides an exponential improvement in speed with
respect to any known classical algorithm. 
The integer factoring problem can be reduced to the problem of 
finding the period of the function $f(k)=a^k$ mod $N$, where $N$ 
is the number to be factorized and $a<N$ is chosen randomly.
The modular exponentiation can be computed efficiently on a 
quantum computer and, starting from the state
$\frac{1}{\sqrt{N}} \sum _{k=0}^{2^n-1} |k\rangle | 0... 0 \rangle$
(the equal superposition of all basis states in the first register can be obtained by 
applying one Hadamard gate for each qubit),
one arrives at
$\frac{1}{\sqrt{N}} \sum _{k=0}^{2^n-1} |k\rangle | f(k) \rangle$.
Notice that there are two quantum registers, the first one stores 
$k$, the second $f(k)$. 
By measuring the second register, one obtains the outcome 
$f(\bar{k})$. Thus, the quantum computer wave function collapses onto
$\frac{1}{\sqrt{m}} \sum_{j=0}^{m-1} |\bar{k} + j r\rangle 
|f(\bar{k})\rangle$,
where $m$ is the number of $k$ values such that $f(k)=f(\bar{k})$,
and $r$ is the period of $f(k)$, that is $f(k)=f(k+r)$.
To determine the period $r$, one has to 
perform the quantum Fourier transform 
of the first register. 
The resulting wave function is peaked around 
integer multiples of $N/r$.
From the measurement of this state, one can extract 
the period $r$. 
It is worth mentioning that
there are cryptographic systems, such as RSA, that are used
extensively today and that are based on the conjecture that
no efficient algorithms exist for solving the prime
factorization problem. Hence Shor's algorithm, if implemented on a large scale
quantum computer, would break the RSA cryptosystem. 

Other quantum algorithms have been developed. In particular, Grover has
shown that quantum computers can also be useful for solving the problem 
of searching for a marked item in an unstructured database of 
$N=2^n$ items. The best one can do with a classical computer is to
go through the database, until one finds the solution. This requires
$O(N)$ operations. In contrast, the same problem can be solved by
a quantum computer in $O(\sqrt{N})$ operations. 
In this case, the gain with respect to classical
computation is quadratic.

\section{Quantum simulation of physical systems}

The simulation of quantum many-body problems on a classical computer is 
a difficult task as the size of the Hilbert space grows exponentially 
with the number of particles. For instance, if one wishes to simulate a chain 
of $n$ spin-$1/2$ particles, the size of the Hilbert space is $2^n$. 
Namely, the state of this system is determined by $2^n$ complex numbers. 
As observed by Feynman in the 1980's, the growth in memory requirement
is only linear on a quantum computer, which is itself a many-body quantum 
system. For example, to simulate $n$ spin-$1/2$ particles one 
only needs $n$ qubits. Therefore, a quantum computer operating with only 
a few tens of qubits can outperform a classical computer. Of course, this is
only true if one can find an efficient quantum algorithm and if one
can efficiently extract useful information from the quantum computer.
Quite interestingly, a quantum computer can outperform a classical computer not only for the
investigation of the properties of many-body quantum systems, but also 
for the study of the quantum and classical dynamics of complex single-particle 
systems.

For a concrete example, consider the quantum-mechanical motion of a
particle in one dimension (the extension to higher dimensions is straightforward). 
It is governed by the Schr{\"o}dinger equation
\begin{equation}
  i \hbar \, \frac{d}{dt} \, \psi(x,t) =
  H \, \psi(x,t) \,,
  \label{1Dsch}
\end{equation}
where the Hamiltonian $H$ is given by
\begin{equation}
  H =
  H_0 + V(x,t) =
  -\frac{\hbar^2}{2m} \, \frac{d^2}{dx^2} + V(x,t) \,.
\end{equation}
The Hamiltonian $H_0=-(\hbar^2\!/2m)\,d^2\!/dx^2$ governs the free motion of
the particle, while $V(x,t)$ is a (possibly time-dependent) one-dimensional potential. To solve
Eq.~(\ref{1Dsch}) on a quantum computer with finite resources (a finite number
of qubits and a finite sequence of quantum gates), one must first of all
discretize the continuous variables $x$ and $t$. If the motion essentially
takes place inside a finite region, say $-d\le{x}\le{d}$, decompose this
region into $2^n$ intervals of length $\Delta=2d/2^n$ and
represent these intervals by means of the Hilbert space of an $n$-qubit
quantum register (this means that the discretization step drops exponentially
with the number of qubits). Hence, the wave function $|\psi(t)\rangle$ is
approximated as follows:
\begin{equation}
  |\tilde{\psi}(t)\rangle =
  \frac1{\mathcal{N}} \sum_{i=0}^{2^{n}-1} \psi(x_i,t) \, |i\rangle \,,
\end{equation}
where
$x_i \equiv -d + \left(i + \tfrac12\right) \Delta$,
$|i\rangle=|i_{n-1}\rangle\otimes\dots\otimes|i_0\rangle$
is a state of the computational basis of the $n$-qubit quantum register and
${\mathcal{N}} \equiv \sqrt{\sum_{i=0}^{2^n-1} |\psi(x_i,t)|^2}$
is a factor that ensures correct normalization of the wave
function. It is intuitive that $|\tilde{\psi}\rangle$ provides a good
approximation to $|\psi\rangle$ when the discretization step $\Delta$
is much smaller than the shortest length scale relevant for the motion of the
system.
The Schr{\"o}dinger equation (\ref{1Dsch}) may be integrated 
by propagating the initial wave function $\psi(x,0)$ 
for each time-step $\epsilon$ as follows:
\begin{equation}
  \psi (x,t + \epsilon) =
  e^{-\frac{i}\hbar [H_0 + V(x,t)] \epsilon} \, \psi(x,t) \,.
\end{equation}
If the time-step $\epsilon$ is small enough, 
it is possible to write the Trotter decomposition
\begin{equation}
  e^{-\frac{i}\hbar [H_0+ V(x,t)] \, \epsilon} \approx
  e^{-\frac{i}\hbar H_0 \, \epsilon}
  e^{-\frac{i}\hbar V(x,t) \, \epsilon} \,,
  \label{Trotter}
\end{equation}
which is exact up to terms of order $\epsilon^2$.
The operator on the right-hand side of
Eq.~(\ref{Trotter}) is still unitary, simpler than that on the
left-hand side, and, in many interesting physical problems,
can be efficiently implemented on a quantum computer. 
Advantage is taken of the fact that the Fourier transform can be
efficiently preformed by a quantum computer. 
One can then write 
the first operator in the
right-hand side of (\ref{Trotter}) as
\begin{equation}
  e^{-\frac{i}\hbar H_0 \, \epsilon} =
  F^{-1} \,
  e^{+\frac{i}\hbar \left(\frac{\hbar^2k^2}{2m}\right) \epsilon} \,
  F \,,
\end{equation}
where $k$ is the variable conjugated to $x$ and $F$ the discrete
Fourier transform.
This represents a transformation
from the $x$-representation to the $k$-representation, in which this operator
is diagonal. Then, using the inverse Fourier transform $F^{-1}$, one returns to
the $x$-representation, in which the operator $\exp(-iV(x,t)\epsilon/\hbar)$ is
diagonal. The wave function $\psi(x,t)$ at time $t=l\epsilon$ is obtained from
the initial wave function $\psi(x,0)$ by applying $l$ times the unitary
operator
\begin{equation}
  F^{-1} \,
  e^{+\frac{i}\hbar \left(\frac{\hbar^2k^2}{2m}\right) \epsilon} \,
  F \,
  e^{-\frac{i}\hbar V(x,t) \, \epsilon}.
\end{equation}
Therefore, simulation of the Schr{\"o}dinger equation is now
reduced to the implementation of the Fourier transform plus diagonal operators
of the form
\begin{equation}
  |x\rangle \to e^{icf(x)} \, |x\rangle \,,
  \label{eifx}
\end{equation}
where $c$ is some real constant. Note that an operator of the form (\ref{eifx})
appears both in the computation of $\exp(-iV(x,t)\epsilon/\hbar)$ and of
$\exp(-iH_0\epsilon/\hbar)$, when this latter operator is written in the
$k$-representation. 
The quantum computation of (\ref{eifx}) is
possible, using an ancillary quantum register $|y\rangle_a$, by means of
the following steps:
\begin{align}
  |0\rangle_a    \otimes |x\rangle
  &\to
  |f(x)\rangle_a \otimes |x\rangle
  \to
  e^{icf(x)} \,
  |f(x)\rangle_a \otimes |x\rangle
  \nonumber
\\
  &\to
  e^{icf(x)} \,
  |0\rangle_a    \otimes |x\rangle
  =
  |0\rangle_a    \otimes  e^{icf(x)} \, |x\rangle \,.
  \label{eifxsteps}
\end{align}
The first step is a standard function evaluation and
may be implemented by means of $O(n2^n)$ elementary
quantum gates. Of course, more efficient implementations
(polynomial in $n$) are possible when the function $f(x)$ has some structure,
as it is the case for the potentials $V(x,t)$ usually considered in 
quantum-mechanical problems. 
The second step in (\ref{eifxsteps}) is the
transformation $|y\rangle_a\to{}e^{icy}|y\rangle_a$ and can be performed in $m$
single-qubit phase-shift gates, $m$ being the number of
qubits in the ancillary register. Indeed, one may write the
binary decomposition of an integer $y\in[0,2^m-1]$ as
$y=\sum_{j=0}^{m-1}y_j2^j$, with $y_j\in\{0,1\}$. Therefore,
\begin{equation}
  \exp(i y) =
  \exp\bigg( \, \sum_{j=0}^{m-1} i c y_j 2^j \bigg) =
  \prod_{j=0}^{m-1} \exp(i c y_j 2^j) \,,
\end{equation}
which is the product of $m$ single-qubit gates, each 
acting non-trivially (differently from identity) only on a
single qubit. The $j$-th gate operates the transformation
$|y_j\rangle_a\to\exp(icy_j2^j)|y_j\rangle_a$, with
$|y_j\rangle_a\in\{|0\rangle,|1\rangle\}$ vectors of the computational basis
for the $j$-th ancillary qubit.
The third step in (\ref{eifxsteps}) is just the reverse of the first 
and may be implemented by the same array of gates as the first
but applied in the reverse order. After this step the ancillary
qubits are returned to their standard configuration $|0\rangle_a$ and
it is therefore possible to use the same ancillary qubits for
every time-step.
Note that the number of ancillary qubits $m$ determines the
resolution in the computation of the diagonal operator (\ref{eifx}).
Indeed, the function $f(x)$ appearing in (\ref{eifx}) is
discretized and can take $2^m$ different values.

An example of an interesting dynamical model that can be
simulated efficiently (and without ancillary qubits) 
on a quantum computer is
the so-called quantum sawtooth map. 
This map represents the dynamics of a periodically driven
system and is derived from the Hamiltonian
\begin{equation}
  H(\theta,I;\tau) =
  \frac{I^2}{2} + V(\theta)
  \sum_{j=-\infty}^{+\infty} \delta(\tau-jT) \,,
  \label{sawham}
\end{equation}
where $(I,\theta)$ are conjugate action-angle variables
($0\leq\theta<2\pi$), with 
the usual quantization rules, $\theta\to\theta$ and
$I\to{}I=-i\partial/\partial\theta$ (set $\hbar=1$)
and $V(\theta)=-k(\theta - \pi)^2/2$.
This Hamiltonian is the sum of two terms,
$H(\theta,I;\tau)=H_0(I)+U(\theta;\tau)$, where $H_0(I)=I^2\!/2$ is
just the kinetic energy of a free rotator (a particle moving on a circle
parametrized by the coordinate $\theta$), while
$U(\theta;\tau) = V(\theta) \sum_j \delta(\tau-jT) $
represents a force acting on the particle 
that is switched on and off instantaneously at time intervals
$T$. Therefore, its is said that the dynamics described by Hamiltonian
(\ref{sawham}) is kicked.
The (quantum) evolution from
time $tT^-$ (prior to the $t$-th kick) to time $(t+1)T^-$
(prior to the $(t+1)$-th kick) is described by a 
unitary operator $U$ acting on the wave function $\psi$:
\begin{align}
  &\psi_{t+1} = U \, \psi_t = U_T U_k \,\psi_t\,;\nonumber\\ 
  \nonumber \\
  &U_T= e^{-i T I^2\!/2} \,,\;\;
  U_k= e^{ik(\theta -\pi)^2\!/2}\,.
  \label{sawquantum}
\end{align}
This map is called the quantum sawtooth map, since the force
$F(\theta)=-dV(\theta)/d\theta=k(\theta-\pi)$ has a sawtooth shape, with a
discontinuity at $\theta=0$. 

In the following, an exponentially efficient quantum algorithm for
simulation of the map (\ref{sawquantum}) is described.
It is based on the forward/backward quantum Fourier transform between action
and angle bases. Such an approach is convenient since the operator $U$
is the product of the two operators
$U_k$ and $U_T$, which are diagonal in the $\theta$ and $I$
representations, respectively. This quantum algorithm requires the following
steps for one map iteration:
\begin{enumerate}
\item
Apply $U_k$ to the wave function $\psi(\theta)$. In order to decompose
the operator $U_k$ into one- and two-qubit gates, we first of all
write $\theta$ in binary notation:
\begin{equation}
  \theta=2\pi\sum_{j=1}^n \alpha_j 2^{-j} \,,
\label{thetaexpand}
\end{equation}
with $\alpha_i\in \{ 0,1 \}$. Here $n$ is the number of qubits, so that the
total number of levels in the quantum sawtooth map is $N=2^n$.
One can insert (\ref{thetaexpand}) into the unitary operator $U_k$,
obtaining the decomposition
\begin{equation}
e^{\imath k(\theta -\pi)^2/2} = \prod_{i,j=1}^n
e^{\imath 2 \pi^2 k( \alpha_i 2^{-i} -
\frac{1}{2n})
(\alpha_j 2^{-j} -
\frac{1}{2n})},
\label{dec}
\end{equation}
which is the product of $n^2$ two-qubit gates, each
acting non-trivially only on the $4$-dimensional
subspace spanned by the qubits $i$ and $j$.
\item
The change from the $\theta$ to the $I$ representation is obtained by means 
of the quantum Fourier transform, which requires 
and $\frac12 n(n+1)$ elementary quantum gates.
\item
In the $I$ representation, the operator $U_T$ has essentially the same
form as the operator $U_k$ in the $\theta$ representation, and therefore
it can be decomposed into $n^2$ two-qubit
gates, similarly to Eq.~(\ref{dec}).
\item
Return to the initial $\theta$ representation by application
of the inverse quantum Fourier transform.
\end{enumerate}
Thus, overall, this quantum algorithm requires $3n^2+n$
gates per map iteration. This number is to be compared with the
$O(n2^n)$ operations required by a classical computer to simulate
one map iteration by means of a fast Fourier transform. Thus,
the quantum simulation of the quantum sawtooth map dynamics is exponentially
faster than any known classical algorithm.
Note that the resources required to the quantum computer to simulate
the evolution of the sawtooth map are only logarithmic in the
Hilbert space dimension $N$.

\begin{figure}
\centering
\includegraphics[scale=0.48]{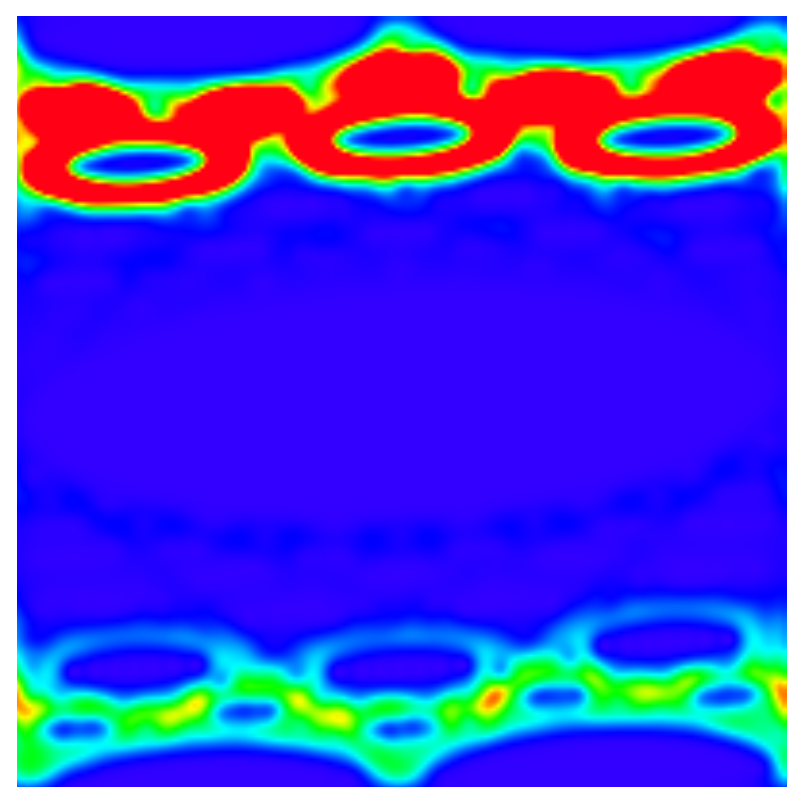}
\includegraphics[scale=0.48]{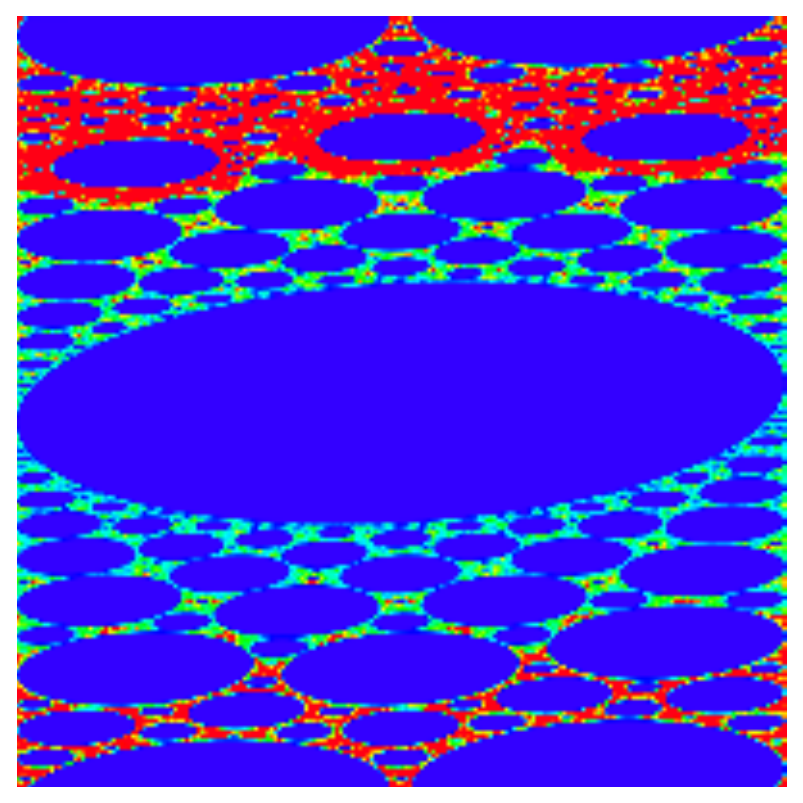}
\caption{Husimi function for the sawtooth map 
for $n=9$ (left) and $n=16$ (right) qubits,
in action angle variables $(I,\theta)$, with
$-N/2 \leq I < N/2$ (vertical axis, $N=2^n$) and
$0\leq \theta < 2\pi$ (horizontal axis), 
averaged in the interval $950\le t \le 1000$,
for $T=2\pi/N$ and $kT=-0.1$.
An action eigenstate,
$|\psi_0\rangle=|m_0\rangle$, with $m_0=[0.38 N]$ is considered as
initial state at time $t=0$.
The color is proportional to the density: blue for zero and
red for maximal density.
}
\label{fig1}
\end{figure}

As an example of the efficiency of this quantum algorithm, 
Fig.~\ref{fig1} shows the Husimi functions, taken after $1000$
map iterations.
It is noted that $n=9$ qubits are sufficient to observe 
the appearance of integrable
islands, while at $n=16$ these islands exhibit a
complex hierarchical structure in the phase space.

However, there is an additional aspect to be taken into account.
Any quantum algorithm has to address the problem of efficiently
extracting useful information from the quantum computer wave function.
Indeed, the result of the simulation of a quantum system is the wave
function of this system, encoded in the $n$ qubits of the quantum
computer. The problem is that, in order to measure all $N=2^n$
wave function coefficients by means of standard polarization
measurements of the $n$ qubits, one has to repeat the quantum
simulation a number of times exponential in the number of qubits.
This procedure would spoil any quantum algorithm, even in the
case, like the present one, 
in which such algorithm could compute the wave function
with an exponential gain with respect to any classical computation.
Nevertheless, there are some important physical questions that can be
answered in an efficient way. 

\begin{figure}[h]
\centering
\vspace{0.5cm}
\includegraphics[scale=0.45]{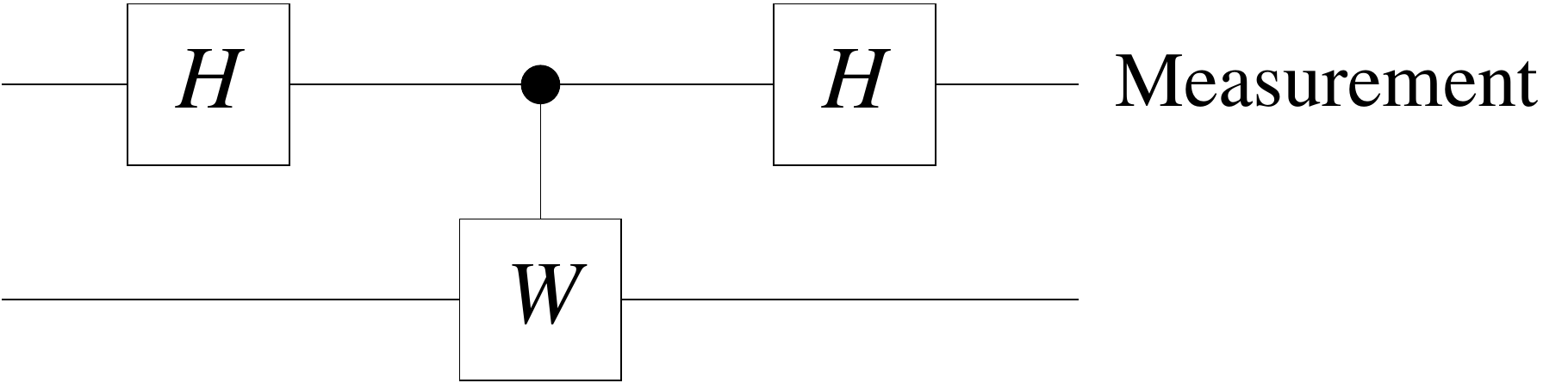}
\caption{Schematic drawing of a quantum circuit implementing a 
Ramsey-type quantum interferometer.
The top line denotes a single ancillary qubit, the bottom 
line a set of $n$ qubits, $H$ the Hadamard gate and $W$ a 
unitary transformation.}
\label{fig2}
\end{figure}

The quantum computation can provide an exponential gain
(with respect to any known classical computation) in problems
that require the simulation of dynamics up to a time $t$ which is
independent of the number of qubits. In this case, provided that
one can extract the relevant information in a number of measurements
polynomial in the number of qubits, one should compare, in 
the example of the quantum sawtooth map,
$O(t(\log{N})^2)$ elementary gates (quantum computation)
with $O(tN\log{N})$ elementary gates (classical computation).
This is, for instance, the case of 
dynamical correlation functions of the form
\begin{equation}
  C_t \equiv
  \langle\psi_0| \, A_t^\dagger \, B_0 \, |\psi_0\rangle =
  \langle\psi_0| \, (U^\dagger)^t \, A_0^\dagger \, U^t \, B_0 \, |\psi_0\rangle
  \,,
  \label{corrfun}
\end{equation}
where $U$ is the time-evolution operator (\ref{sawquantum}) for the 
quantum sawtooth
map. Similarly, one can efficiently compute the fidelity of quantum
motion,
which is a quantity of central interest in the study of the stability
of quantum motion under perturbations.
The fidelity $f(t)$ (also called
the Loschmidt echo), measures the accuracy with which a quantum
state can be recovered by inverting, at time $t$, the dynamics with a 
perturbed Hamiltonian. It is defined as
\begin{equation}
  f(t) = \langle\psi| \, (U^\dagger_\epsilon)^t \, U^t \, |\psi\rangle 
  = \langle\psi| \, e^{iH_\epsilon t} \, e^{-iHt} \, |\psi\rangle 
\,.
\end{equation}
Here the wave vector $|\psi\rangle$ evolves forward in time with 
Hamiltonian $H$ up to time $t$ and then evolves
backward in time with a perturbed Hamiltonian $H_\epsilon$. 
If the evolution operators $U$ and $U_\epsilon$ 
can be simulated efficiently on a quantum computer, as is 
the case in many physically interesting situations, then the fidelity of 
quantum motion can be evaluated with exponential speed up with respect 
to known classical computations. 
As shown in Fig.~\ref{fig2}, it is possible to measure the fidelity
by means of a Ramsey-type quantum interferometer method. 
A single ancillary qubit is needed, initially prepared in the state
$|0\rangle$, while the input state for the other $n$ qubits is 
a given initial state $|\psi_0\rangle$ for the quantum sawtooth map. 
Two Hadamard gates are applied to the ancillary qubit, and in between
these operations a controlled-$W$ operation is applied ($W$ is a 
unitary operator), namely $W$ is applied to the other $n$ qubits 
only if the ancillary qubit is in its $|1\rangle$ state. 
As a result, one obtains the following 
final overall state for the $n+1$ qubits:
\begin{equation}
\frac{1}{2}\,
[(|0\rangle +|1\rangle) |\psi_0\rangle 
+ (|0\rangle -|1\rangle) W|\psi_0\rangle].
\end{equation}
If $W=(U_\epsilon^\dagger)^t U^t$, then one can derive the fidelity 
from polarization measurements of the ancillary qubit. One obtains
\begin{equation}
f(t)= \langle \sigma_z \rangle^2+ 
\langle \sigma_y \rangle^2,
\end{equation}
where 
$\langle \sigma_z\rangle$ and $\langle \sigma_y\rangle$ are the
expectation values of the Pauli operators
$\sigma_z$ and 
$\sigma_y$. 
Provided that the quantum algorithm implementing $U$ is efficient,
as it is the case for the quantum sawtooth map, the fidelity 
can then be computed efficiently.

\section{Simulating complex dynamics on actual quantum hardware}

Present-day quantum computers, whether they are based on superconducting qubits or on trapped ions, 
suffer from significant decoherence and the effects of various noise sources.
Therefore, achieving the quantum advantage in practically relevant problems 
such as chemical reactions, new materials design, or biological processes, is an imposing task.
Note that quantum advantage is achieved when a quantum computer can solve a problem that no classical 
computer can solve in a feasible amount of time. The progress of currently available quantum processors
can nevertheless be benchmarked by simulating complex dynamics. 

An illustrative example is again provided by the quantum sawtooth map. The classical limit of such map is chaotic 
when $kT<-4$ or $kT>0$. 
Although the sawtooth map is a deterministic system, in the chaotic regime the motion along the  
action direction is in practice indistinguishable from a random walk, with diffusion in the
action variable. 
If one considers a classical ensemble of trajectories with fixed initial
action $m_0$ and random initial angle $\theta$, the second moment of the action distribution 
grows linearly with the number $t$ of map iterations, $\langle (\Delta I)^2 \rangle \approx D(k) t$,
with a diffusion coefficient $D$ dependent on $k$.  
The quantum sawtooth map, in agreement with the correspondence principle, initially
exhibits diffusive behavior, with the classical diffusion coefficient $D$. However, after a break time 
$t^\star$, quantum interference leads to suppression of diffusion. 
For $t>t^\star$, the quantum distribution reaches a steady state which decays exponentially over the action eigenbasis:
\begin{equation}
  W_m \equiv \big| \langle m | \psi \rangle \big|^2 \approx
  \frac1{\ell} \, \exp\!\left[ -\frac{2|m-m_0|}{\ell} \right] ,
  \label{expdecay}
\end{equation}
where the index $m$ singles out the action eigenstates ($I|m\rangle=m|m\rangle$),
the system is initially prepared in the eigenstate $|m_0\rangle$, and 
$\ell$ is known as the localization length of the system.
Therefore, for
$t>t^\star$ only 
\begin{equation}
\sqrt{\langle(\Delta I)^2\rangle}\approx \sqrt{D t^\star}\approx \ell
\label{siberia1}
\end{equation}
levels are populated.
This phenomenon, known as dynamical localization, is due to quantum interference effects, suppressing the
underlying classical diffusion process after a time $t^\star\approx \ell\approx D$.

\begin{figure}[t]
    \centering
    \includegraphics[width=8cm]{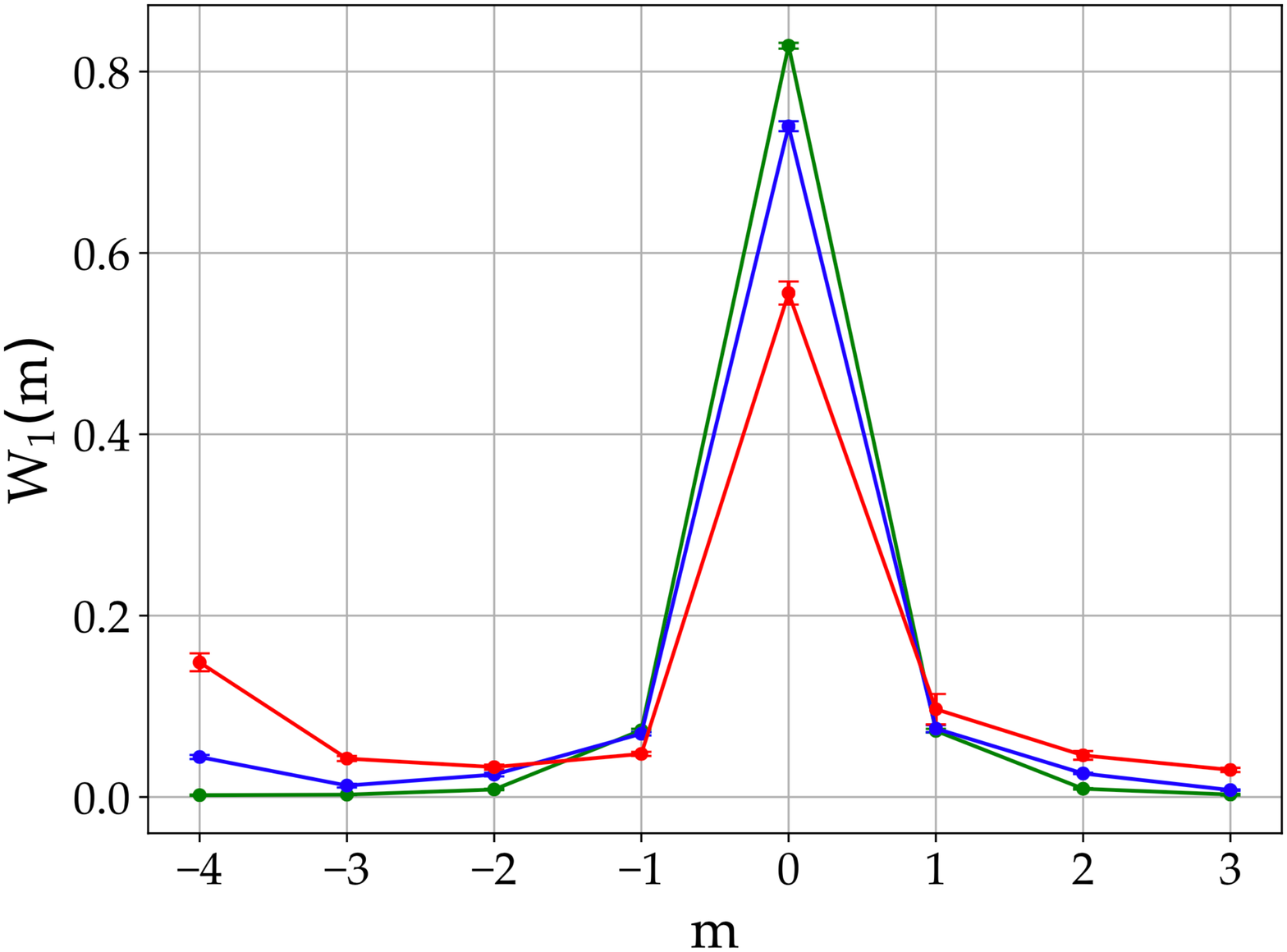}
    \caption{Dynamical localization in the quantum sawtooth map with $n=3$ qubits, 
    $kT=1.5$, $k\approx 0.273$.
    Data from the IBM quantum processors {\it lima} (red) are obtained after averaging over 10 repetitions of 8192 experimental runs,
    and compared with the Qiskit simulator (blue) and the noiseless simulation (green).}
    \label{fig:distribution}
\end{figure}

Fig.~\ref{fig:distribution}  shows the results of a dynamical localization experiment with $n=3$ qubits 
on a real and freely available IBM quantum processor, with superconducting qubits, 
remotely accessed through cloud quantum programming~\cite{localization}. 
The initial condition is peaked in action,
$\psi_0(m)=\langle m |\psi_0\rangle=\delta_{m,m_0}$, with $m_0=0$. The quantum algorithm for the
sawtooth map allows one to compute the wave vector $\psi_t(m)$ as a function
of the number of map steps, and then the probability distribution 
$W_t(m)=|\psi_t(m)|^2$. In the figure $k\approx 0.273 < 1$, so that the 
distribution is already localized after a single map step. On the other hand, 
here $k T=1.5$, corresponding to diffusive, chaotic behavior for the underlying classical 
dynamics.
In Fig.~\ref{fig:distribution}  the 
ideal, noiseless distribution after $t=1$ map step is compared  with the 
results of the real quantum hardware and with a simulator 
(Qiskit, provided by IBM), which takes into account a few relevant noise sources, 
modeling in particular dephasing, relaxation, and readout errors.  
The results show that the quantum hardware exhibits a localization peak, 
which emerges from quantum interference. Note that
the quantum algorithm performs forward-backward Fourier transform, thus exploring the entire Hilbert space of the quantum register
in a complex multiple-path interferometer that leads to wave-function localization. As such, dynamical localization is a very fragile quantum phenomenon, extremely sensitive to noise.
The height of the peak, 
is significantly smaller than 
the noiseless value
and the prediction of the Qiskit simulator.
These results show that the Qiskit
simulator underestimate some of the relevant noise channels,
such as fluctuations of the qubit quality parameters between calibrations 
of the quantum computer, memory effects, and cross-talks between qubits.

The presence of these noise channels also shows the imposing difficulties 
in scaling quantum algorithms to a large number of qubits and of quantum gates. 
On the other hand, striking progress has been reported in recent years, 
quantified for instance by the quantum volume $V_Q$, a single number meant to encapsulate the quantum computer performance,
including number of available qubits and number of quantum gates that can be reliably implemented, before errors dominate~\cite{qvolume}.
The quantum volume is defined as
\begin{equation}
\log_2 V_Q = {\arg\max}_{\kappa\le n}\{\min[\kappa,d(\kappa)]\},
\end{equation}
where $n$ is the number of qubits in the quantum computer, and $d(\kappa)=1/(\kappa\epsilon_{\rm eff}(\kappa))$,
known as circuit depth, is determined by an effective error rate $\epsilon_{\rm eff}(\kappa)$ for a subset 
of $\kappa\le n$ qubits, on which sequences of random two-qubit unitaries are implemented.
From January, 2020 to December, 2021, the reported values of $V_Q$ have increased from $V_Q=32$ 
to $V_Q=2048$ (for a comparison, data from Fig.~\ref{fig:distribution} have been obtained with a machine
with quantum volume $V_Q=8$).

\section{Outlook}

A few significant examples have been discussed showing the 
capabilities of a quantum computer in the simulation of
complex physical systems. A quantum computer with a few
tens of qubits and a long enough decoherence time to allow the implementation of  
a large number of quantum gates, would outperform a 
classical computer in this kind of problems. 

Any practical implementation of a quantum computer has
to face errors, due to the inevitable coupling of the computer
to the surrounding environment or to  
imperfections in the quantum computer hardware.
The first kind of error is known as decoherence and
is a threat to the actual implementation of any quantum computation. 
More generally,
decoherence theory has a fundamental interest beyond quantum information
science, since it provides explanations for the emergence of classicality
in a world governed by the laws of quantum mechanics~\cite{zurek}. The core of the
problem is the superposition  principle, according to which any
superposition of quantum states is an acceptable quantum state.
This entails consequences that are absurd according
to classical intuition, like the superposition of ``cat alive'' and
``cat dead'' that is considered in the Schr\"odinger's cat paradox.
The interaction with the environment can destroy the coherence
between the states appearing in a superposition (for instance,
the ``cat alive'' and ``cat dead'' states).
Therefore, decoherence invalidates the
quantum superposition principle, which is at the heart of the
power of quantum algorithms. 
The presence of device imperfections, although not leading to any 
decoherence, also hinders the implementation of any quantum computational 
task, introducing errors.
Therefore, decoherence and imperfection effects
appear to be the ultimate obstacle to the realization of a 
large-scale quantum computer.

Note that a quantum computer is not necessarily required 
for implementing quantum simulation. Simpler quantum devices, called 
(analog) quantum simulators can mimic the evolution of other quantum systems 
in an analog manner. Such simulators are problem-specific quantum machines,
namely controllable quantum systems used to simulate other quantum systems~\cite{nori}.

At present (2022) it is not clear if and when a useful quantum computer,
capable of outperforming existing classical computers in important computational tasks, will be built. 
In order to perform coherent controlled evolution of a many-qubit system, one needs to take into account the problem of decoherence,
and therefore large-scale quantum computers appear unrealistic with present technology. 
On the other hand, progress in the field has been huge 
in recent years. Moreover, we should bear in mind that technological breakthroughs (such as the transistor was for the classical computer) are always possible and that no fundamental objections have been found against the possibility of building a quantum computer.

\acknowledgments{We acknowledge use of the IBM
Quantum Experience for Fig. 3 of this work. 
The views expressed are those of the authors and do not reflect the official policy or position of IBM company or the IBM-Q team.}

\bibliographystyle{prsty}

\begin{thebibliography}{9}


\bibitem{qcbook} G. Benenti, G. Casati, D. Rossini, and G. Strini,
{\it Principles of Quantum Computation and Information} (A comprehensive textbook)
(World Scientific, Singapore, 2019).

\bibitem{localization}
A. Pizzamiglio, S. Y. Chang, M. Bondani, S. Montangero,
D. Gerace, and G. Benenti,
{\it Dynamical localization simulated on actual quantum hardware},
Entropy {\bf 23}, 654 (2021).

\bibitem{qvolume}
A. W. Cross, L. S. Bishop, S. Sheldon, P. D. Nation, and J. M. Gambetta,
{\it Validating quantum computers using randomized model circuits},
Phys. Rev. A \textbf{100}, 032328 (2019).

\bibitem{zurek} W.H. Zurek,
{\it Decoherence, einselection, and the quantum origins of the classical},
Rev. Mod. Phys. \textbf{75}, 715 (2003).

\bibitem{nori}
I. M. Georgescu, A. Ashhab, and F. Nori, 
{\it Quantum simulation},
Rev. Mod. Phys. \textbf{86}, 153 (2014).

\end{thebibliography}

\section{Keywords}

quantum computers

quantum algorithms

quantum simulation

quantum logic

superposition principle

entanglement

quantum measurements

information extraction

dynamical systems

\end{document}